\documentclass[fleqn,12pt,twoside]{article}
\usepackage{amsmath}
\usepackage{espcrc1}

\usepackage{txfonts}

\usepackage{graphicx}

\newcommand{\chemical}[2]{\ensuremath{{}^{#1}\mathrm{#2}}}

\newcommand{\fm}{\mathrm{fm}}
\newcommand{\MeV}{\mathrm{MeV}}

\renewcommand{\vec}[1]{\boldsymbol{{#1}}}

\newcommand{\ket}[1]{\big| \,{#1}\, \big> }

\newcommand{\braket}[2]{\big< \,{#1}\, \big| \,{#2}\, \big> }

\newcommand{\matrixe}[3]{\big< \,{#1}\, \big| \,{#2}\, 
\big| \,{#3}\, \big> }

\title{Cluster structures within Fermionic Molecular Dynamics}

\author{T. Neff, H. Feldmeier\\[2ex]
  Gesellschaft f\"ur Schwerionenforschung mbH, \\
    Planckstra\ss e 1, 64291 Darmstadt, Germany}

\begin{document}

\maketitle

\begin{abstract}
  The many-body states in an extended Fermionic Molecular Dynamics
  approach are flexible enough to allow the description of nuclei with
  shell model nature as well as nuclei with cluster and halo
  structures. Different many-body configurations are obtained by
  minimizing the energy under constraints on collective
  variables like radius, dipole, quadrupole and octupole deformations.
  In the sense of the Generator Coordinate Method we perform variation
  after projection and multiconfiguration calculations. The same
  effective interaction derived from realistic interactions by means
  of the Unitary Correlation Operator Method is used for all nuclei.
  Aspects of the shell model and cluster nature of the ground and
  excited states of \chemical{12}{C} are discussed. To understand
  energies and radii of neutron-rich He isotopes the soft-dipole mode
  is found to be important.
\end{abstract}

\section{FERMIONIC MOLECULAR DYNAMICS}

In the Fermionic Molecular Dynamics (FMD) model \cite{fmd00} the A-body
state is given as a Slater determinant $\ket{Q}$ of
single-particle states $\ket{q_i}$
\begin{equation}
  \label{eq:fmdslaterdet}
  \ket{Q} = \mathcal{A} \biggl\{ \ket{q_1} \otimes \ldots \otimes
    \ket{q_A} \biggr\} \: .
\end{equation}
The single-particle wave functions are described by Gaussian wave
packets localized in phase-space
\begin{equation}
  \braket{\vec{x}}{q}=\sum_{i} c_i \exp \biggl\{ -
       \frac{(\vec{x} -\vec{b}_i)^2}{2a_i} \biggr\}
        \ket{\chi_i} \otimes \ket{\xi} \: .
\end{equation}
In contrast to the AMD approach \cite{amd01} in the FMD model the
width parameter $a$ is treated as a complex
variational parameter that can be different for each single-particle
state. The spin orientation given by the spinor $\ket{\chi}$ is also
treated as a variational parameter.  For light nuclei the description
can be improved by using a superposition of two Gaussian wave packets.


The FMD many-particle state is determined by minimizing the intrinsic
energy
\begin{equation}
  E\left[\ket{Q}\right] = \frac{\matrixe{Q}{H_{\mathit{eff}}-T_{\mathit{cm}}}{Q}}{\braket{Q}{Q}}
\end{equation}
with respect to the parameters of all single-particle states. This
corresponds to
a Hartree-Fock calculation using the FMD single-particle basis. To
improve upon this mean-field result we restore the symmetries of the
Hamiltonian and perform a parity and angular momentum projection. This
defines the projection after variation (PAV) result.

The effects of the projection can be quite large and a variation after
projection calculation (VAP) should be done \cite{amd01}. But as a
full VAP calculation is very expensive we perform VAP
calculations in the spirit of the Generator Coordinate Method. The
intrinsic energy is minimized under constraints on collective
coordinates like radius, dipole, quadrupole or octupole moments
obtaining a set of many-body configurations. The
VAP minimum is then obtained by minimizing the projected energies of
these configurations as a function of the constraints. These
configurations can then also be used in a multiconfiguration
calculation (MC) where the projected Hamiltonian is diagonalized
within the set of these configurations. This provides us with a well
defined procedure to study the properties of the ground and the
excited states.

\section{EFFECTIVE INTERACTION}

\begin{figure}[h]
  \includegraphics[angle=0,width=0.24\textwidth]{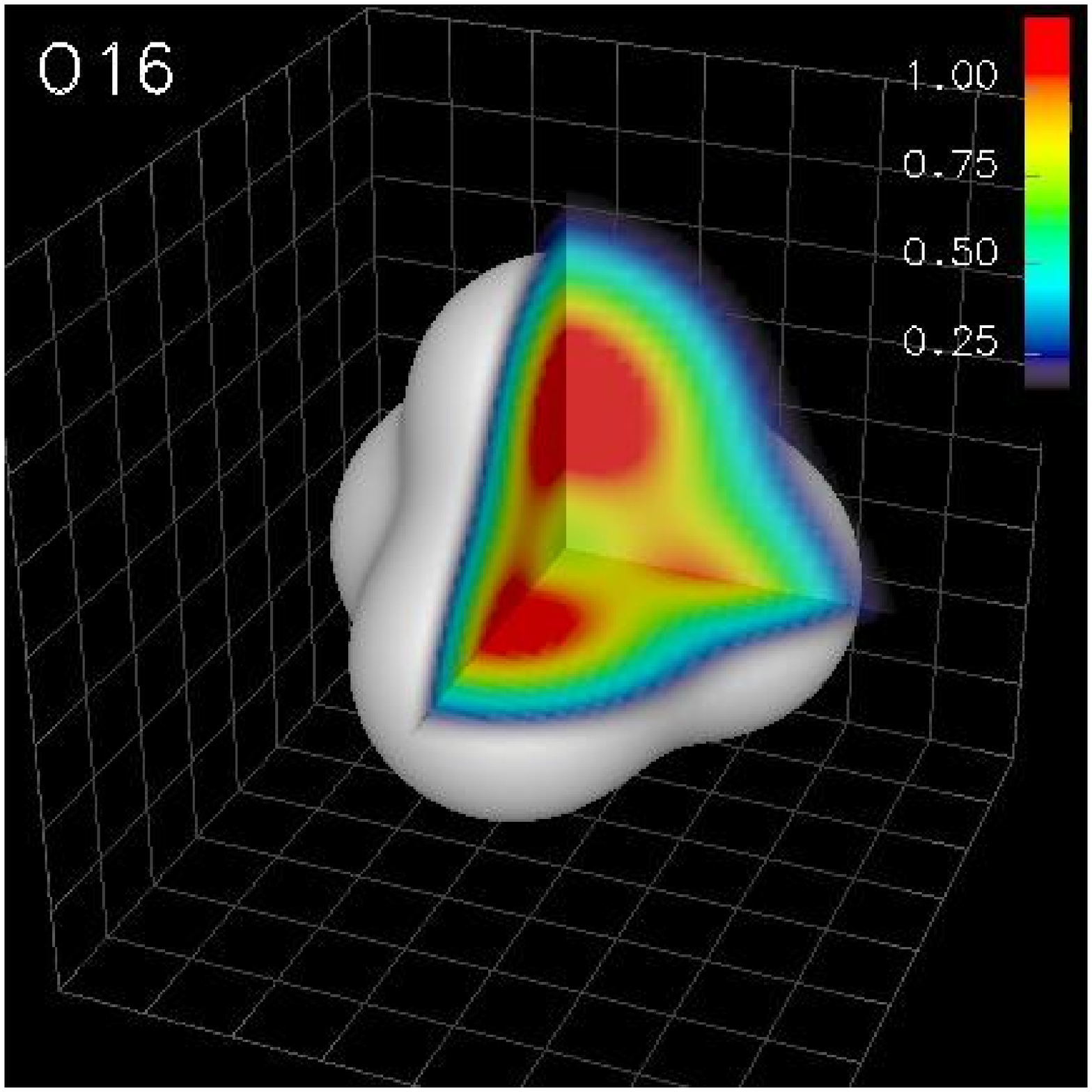}\hfil
  \includegraphics[angle=0,width=0.24\textwidth]{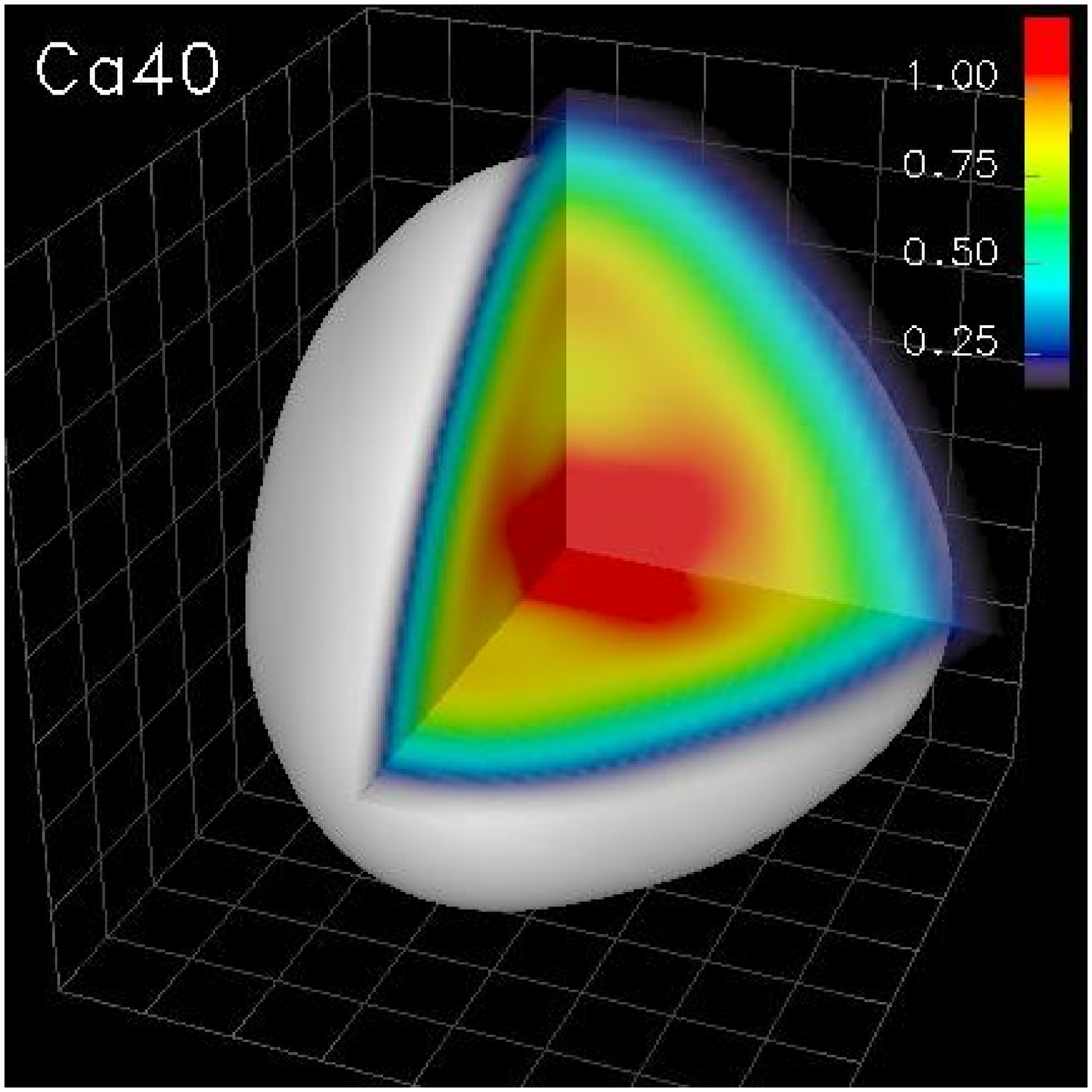}\hfil
  \includegraphics[angle=0,width=0.24\textwidth]{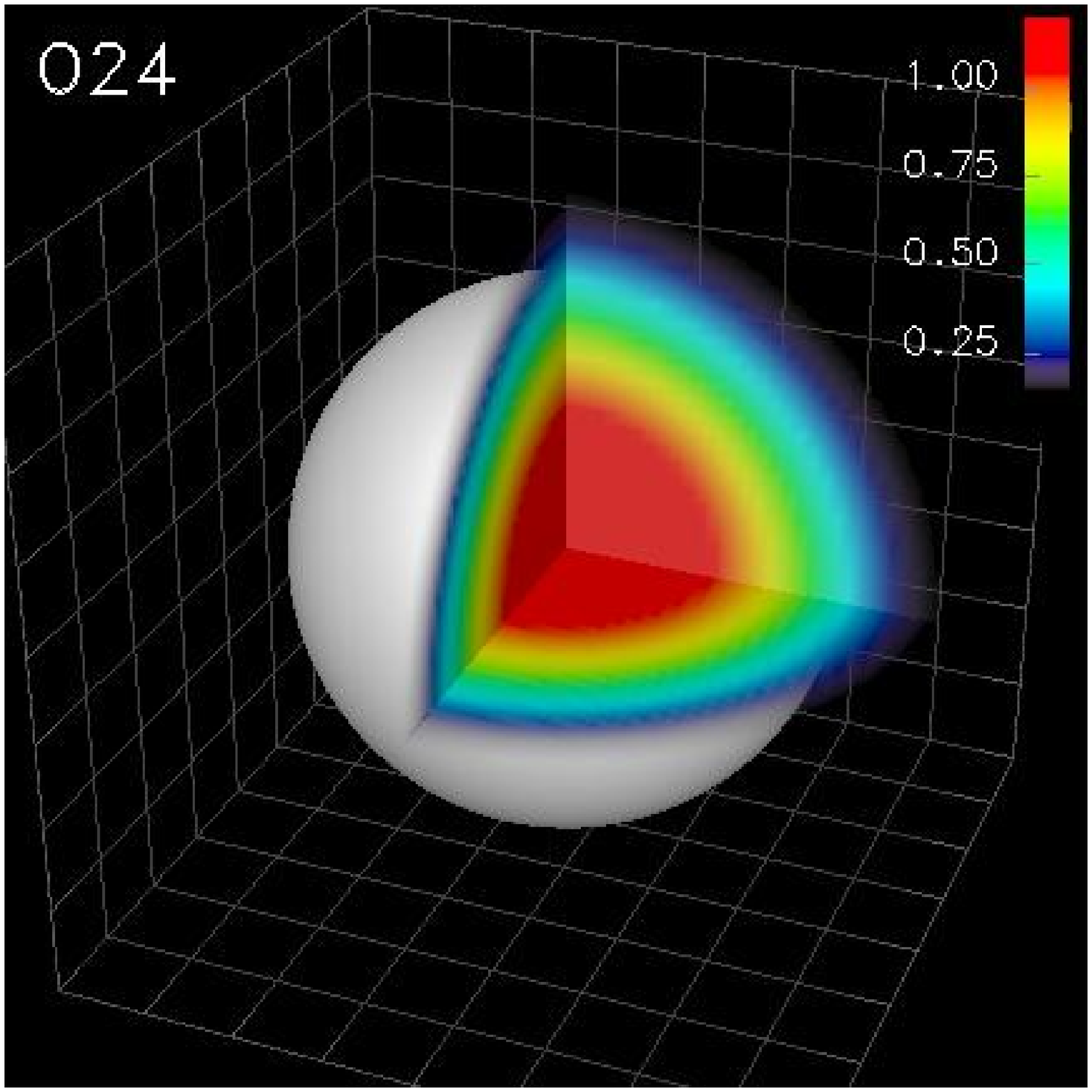}\hfil
  \includegraphics[angle=0,width=0.24\textwidth]{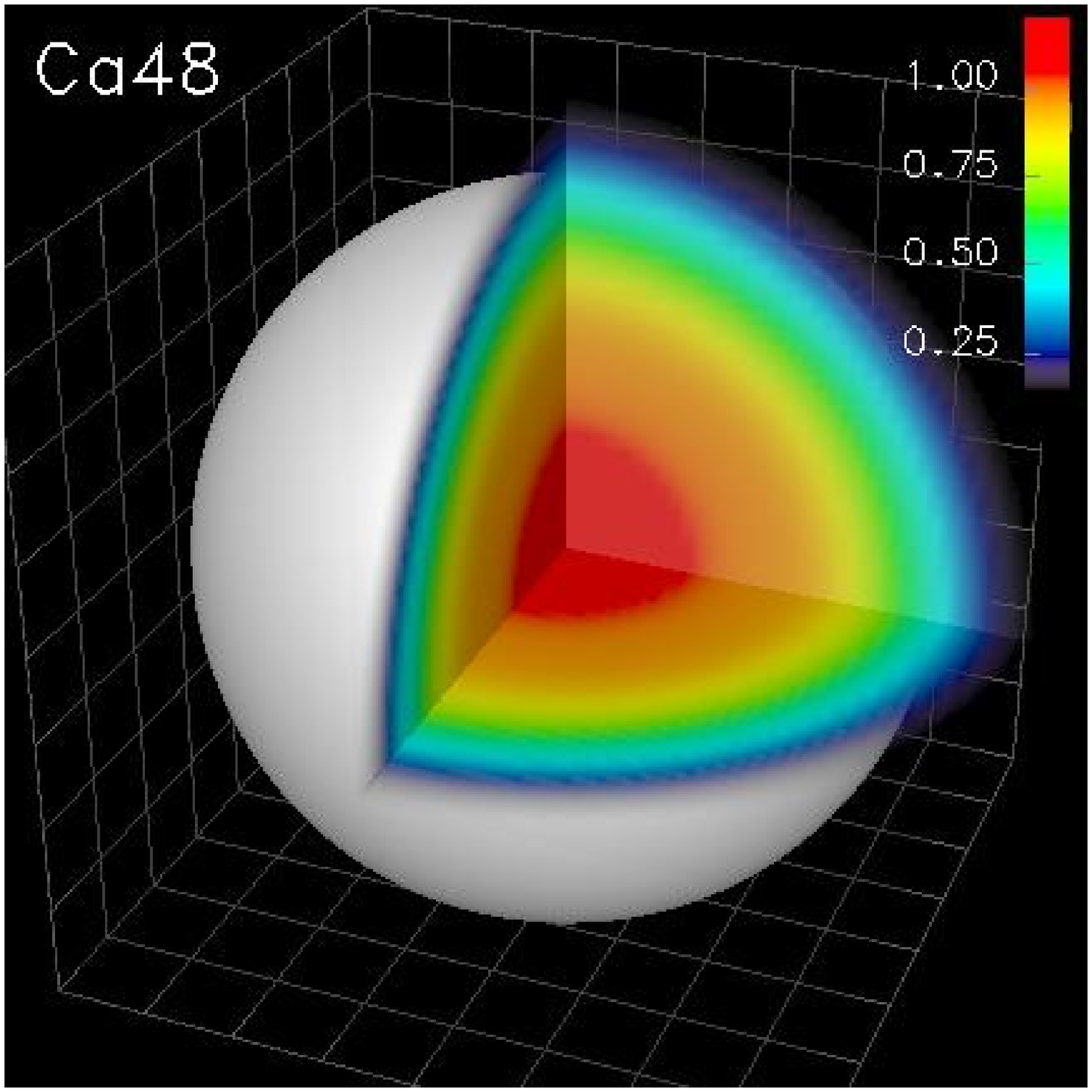}\hfil
  \vspace{-4ex}
  \caption{The effective interaction is fitted to tetrahedral
    $\chemical{16}{O}$ and $\chemical{40}{Ca}$ and spherical
    $\chemical{24}{O}$ and $\chemical{48}{Ca}$ configurations.}
  \label{fig:fitnuclei}
\end{figure}

We use an effective interaction that is derived from the realistic
Bonn or Argonne interactions by means of the Unitary Correlation
Operator Method (UCOM) \cite{ucom98,ucom03}. The correlated interaction
includes the short-range radial and tensor correlations induced by
the repulsive core and the tensor force. The correlated interaction no
longer connects to high momenta and can thus be used directly within
product-state model spaces. The effects of three-body correlations and
genuine three-body forces are at this stage simulated by an additional
two-body force that contains momentum-dependence and spin-orbit
forces. This correction term is adjusted to reproduce the binding
energies and radii of $\chemical{4}{He}$, $\chemical{16}{O}$,
$\chemical{40}{Ca}$, $\chemical{24}{O}$ and $\chemical{48}{Ca}$. Here
we find for $\chemical{16}{O}$ and $\chemical{40}{Ca}$ that after
angular momentum projection the tetrahedral configurations (see
Figure~\ref{fig:fitnuclei}) with $\alpha$-clusters are favored
energetically versus the spherical shell model configurations by about
5~MeV. Altogether we need a 15\% correction to the \emph{ab-initio}
correlated two-body interaction.

\vspace{-3ex}

\vspace{-2ex}




\section{GROUND AND EXCITED STATES OF ${}^{12}$C}

\begin{figure}[t]
\begin{minipage}[t]{0.33\textwidth}
  \setlength{\unitlength}{\textwidth}
  \begin{picture}(1,1.68)
    \put(0,1.12){\makebox(1,0.56){
        \includegraphics[angle=0,width=.5\unitlength]{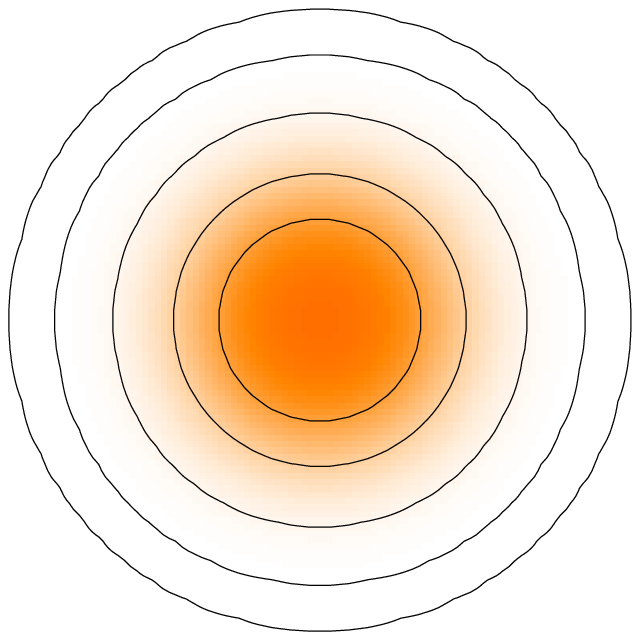}
        \includegraphics[angle=0,width=.5\unitlength]{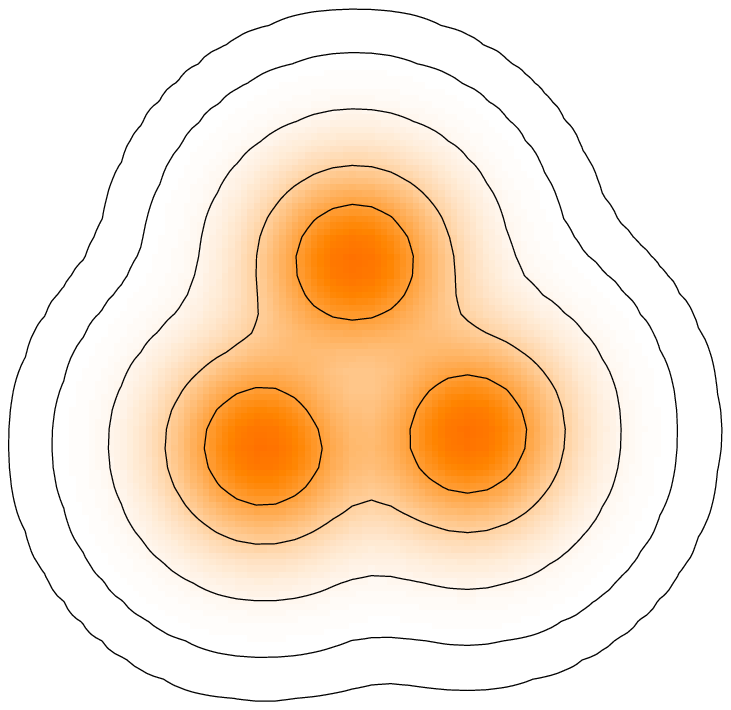}
      }}
    \put(0.11,1.12){V/PAV} \put(0.55,1.12){VAP $\alpha$-cluster}
    \put(0,0.56){\makebox(1,0.56){
        \includegraphics[angle=0,width=.5\unitlength]{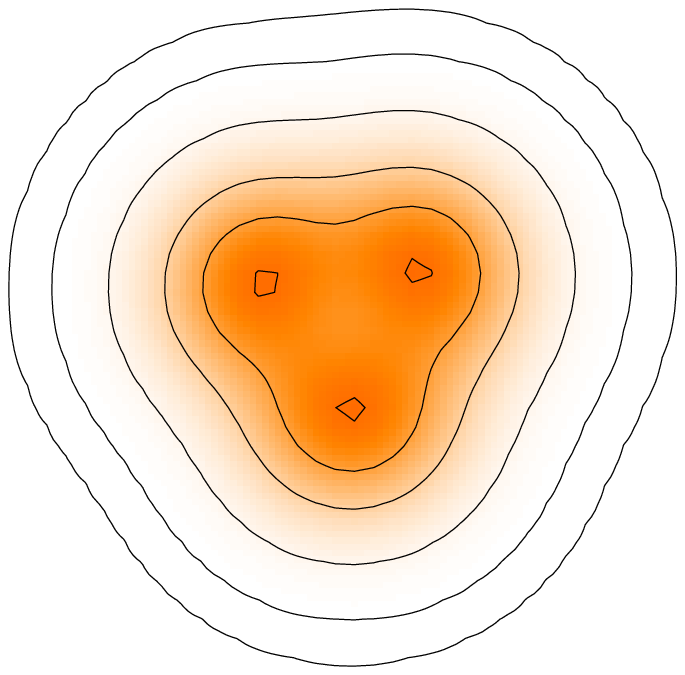}
        \includegraphics[angle=0,width=.5\unitlength]{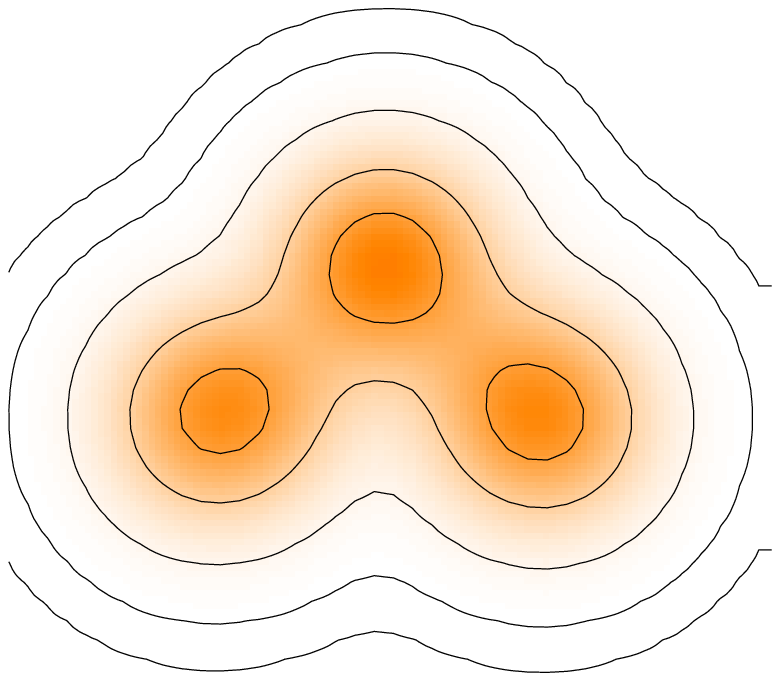}
      } }
    \put(0.16,0.55){VAP} \put(0.68,0.55){`` $3_1^-$ ''}
    \put(0,0.0){\makebox(1,0.56){
        
        \includegraphics[angle=0,width=.5\unitlength]{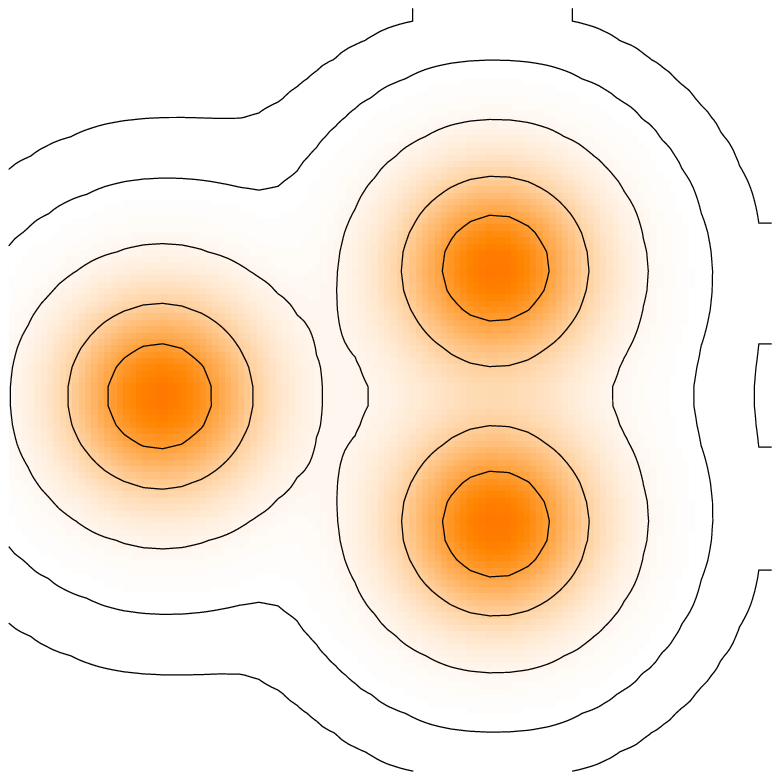}
        \includegraphics[angle=0,width=.5\unitlength]{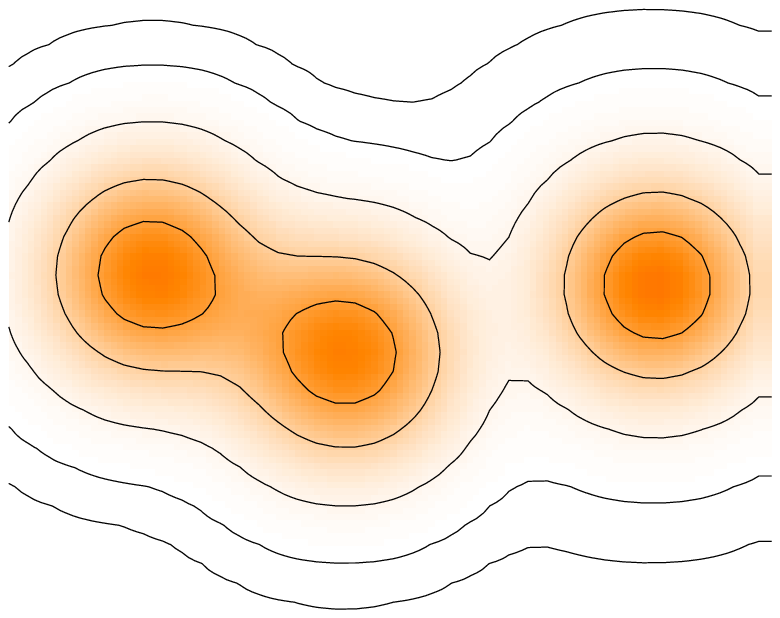}
      } }
    \put(0.2,-0.05){`` $0_2^+$ ''} \put(0.68,-0.05){`` $0_3^+$ ''}
  \end{picture}

\end{minipage}
\hspace{\fill}
\begin{minipage}[b]{0.65\textwidth}
  \includegraphics[angle=90,width=\textwidth]{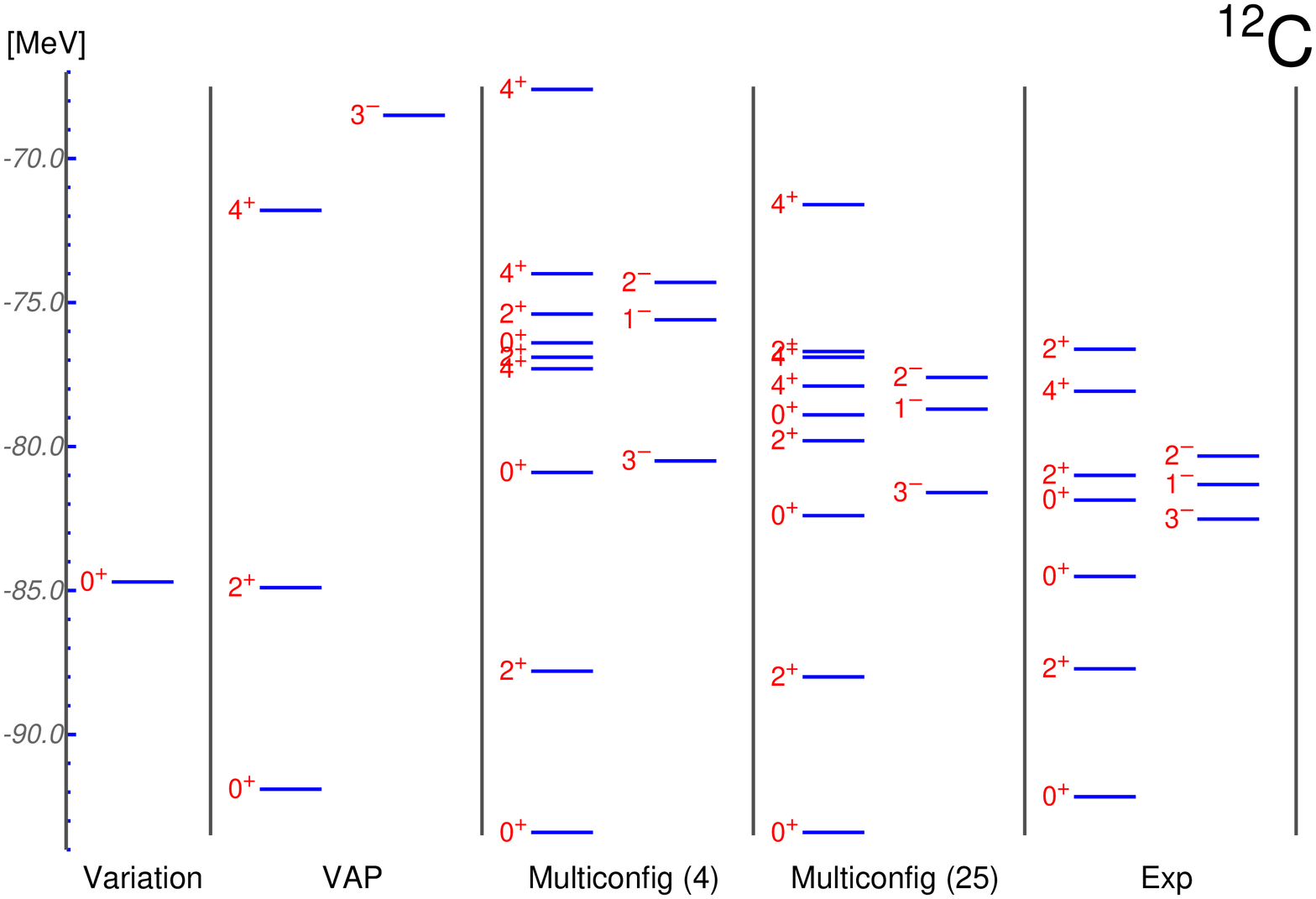}
  \vspace{-5ex}
  \caption{Intrinsic shapes used in the \chemical{12}{C} calculations
    (l.h.s.). Calculated and experimental spectrum of \chemical{12}{C} (r.h.s.)}
  \label{fig:c12}
\end{minipage}
\end{figure}

\begin{table}[b]
  \renewcommand{\arraystretch}{1.2}
  \caption{Binding energies, charge radii and $BE(2)$-values of $^{12}$C.}
  \begin{tabular}{@{}lrrr@{}} \hline 
        & $E_b \; [\MeV]$ & $r_{\mathit{charge}}\; [\fm]$ & $B(E2) \;
        [e^2 \fm^4]$ \\ \hline
        V/PAV & 84.7 & 2.33 & - \\
        VAP $\alpha$-cluster & 80.4 & 2.66 & 56.3 \\
        VAP & 91.9 & 2.38 & 24.7 \\
        Multiconfig(4) & 93.4 & 2.50 & 40.0 \\ \hline
        Exp & 92.2 & 2.47 & $39.7 \pm 3.3$\\ \hline
  \end{tabular}
\label{tab:c12}
\end{table}

The question of the clustering in \chemical{12}{C} has a long history
and has been studied in the AMD framework \cite{enyo98,itagaki03}.
Our effective interaction is fitted to binding energies and radii of
\chemical{4}{He} and \chemical{16}{O} but has not been adjusted to
\chemical{12}{C}. We perform the FMD variation and obtain the
$(1s_{1/2})^4(1p_{3/2})^{12}$ shell model configuration (V), displayed
in the left part of Figure~\ref{fig:c12}. As it is spherical the
angular momentum projection (PAV) does not affect the result.  The
binding energy obtained with this configuration is 7.5~MeV too low
compared to the experimental value (see Table~\ref{tab:c12}). The
obtained radius is too small. We can also restrict the intrinsic
state to pure $\alpha$-cluster configurations. In these configurations
the spin-orbit interaction does not contribute and already in the
mean-field calculation a triangular configuration is preferred. In a
VAP calculation where the positions and the widths of the $\alpha$'s
are the generator coordinates we obtain 4.3~MeV less binding than with
the shell model configuration and the radius is too large for the
pure $\alpha$-cluster configuration. If we perform a variation after
projection calculation with the radius and the octupole moment as
generator coordinates we obtain the triangular configuration with
polarized $\alpha$-clusters denoted by VAP on the l.h.s. of
Figure~\ref{fig:c12}. It represents an intermediate between the VAP
$\alpha$-cluster and the shell model (V/PAV) shape.
The description of this polarization is
strongly affected by using two Gaussians per single-particle state. We
obtain 4.2~MeV more binding compared to a VAP calculation with only
one Gaussian per single-particle state. In case of the shell-model
configuration the energy gain is only 1.5~MeV. The radius obtained in
the VAP calculation is a little bit too small when compared to
experiment. For the intrinsically deformed VAP configurations we can
project out not only a $0^+$ but also $2^+$, $4^+$ and $3^-$ states
(as shown in the VAP column of the spectrum in Figure~\ref{fig:c12})
and can calculate the $B(E2)$ value for the $0^+ \rightarrow 2^+$
transition using bare charges. The $B(E2)$ value obtained for the VAP
configuration is too small compared to the experimental value.

To improve the description we perform multiconfiguration calculations.
For the additional configurations quadrupole and octupole moments are
used as generator coordinates. In Figure~\ref{fig:c12} the
configurations giving the biggest contributions for the first $3^-$
and the second and third $0^+$ states are shown. In a
multiconfiguration calculation containing the VAP and these three
intrinsic states the binding energy of the ground state is increased
by 1.5~MeV.  For the electromagnetic properties the effect of admixing
these less compact configurations is more pronounced. The radius is
increased and the $B(E2)$ value is now in agreement with the
experimental value. The spectrum obtained with these four intrinsic
states is shown in the Multiconfig(4) column of Figure~\ref{fig:c12}.
Using 25 configurations the energies for the ground state and the
lowest excited states change only slightly. For the higher excited
states the effect is much bigger. Comparing to the experimental
spectrum our results for the excited $0^+$ and $2^+$ states are
already quite reasonable. Whereas the second $0^+$ state seems to be
mainly built of three $\alpha$ configurations, the third $0^+$ appears
to be mainly \chemical{8}{Be} plus $\alpha$. A detailed study of these
$0^+$ and also $2^+$ states will be done in the future.  We expect
that even more extended configurations will have to be considered,
taking into account the proposed Bose condensed nature of these states
around the three $\alpha$ threshold \cite{horiuchi03}.

\section{HELIUM ISOTOPES}

\begin{table}[b]
  \caption{Experimental and calculated binding energies and radii for
    the He isotopes.}
  \renewcommand{\arraystretch}{1.2}
  \begin{tabular}{@{}llrrrrr@{}} \hline
    & & \phantom{xxxxxx} ${}^4$He & \phantom{xxxxxx} ${}^5$He &
    \phantom{xxxxxx} ${}^6$He & \phantom{xxxxxx} ${}^7$He &
    \phantom{xxxxxx} ${}^8$He \\ \hline
    PAV & $E_b$ [MeV]           & 28.3 & 25.4 & 26.3 & 26.5 & 29.8\\
    VAP & $E_b$ [MeV]           & 28.3 & 26.8 & 27.7 & 27.7 & 31.1\\
    Multiconf & $E_b$ [MeV]            & 28.4 & 27.4 & 29.1 & 28.7 & 31.7 \\
    Exp & $E_b$ [MeV]           & 28.3 & 27.4 & 29.3 & 28.8 & 31.4\\ \hline
    Multiconf & $r_\textit{mat}$ [fm]  & 1.45 & 2.74 & 2.42 & 2.62 & 2.53\\
    Exp & $r_\textit{mat}$ [fm] &  &      & 2.48 $\pm$ 0.03 & & 2.52 $\pm$ 0.03\\ \hline
    Multiconf & $r_\textit{ch}$ [fm]   & 1.69 & 2.10 & 2.02 & 2.07 &
    2.03\\
    Exp & $r_\textit{ch}$ [fm]   & 1.68 & & & & \\ \hline
  \end{tabular}
  \label{tab:heisotopes}
\end{table}

\begin{figure}[tb]
  \includegraphics[angle=0,width=0.48\textwidth]{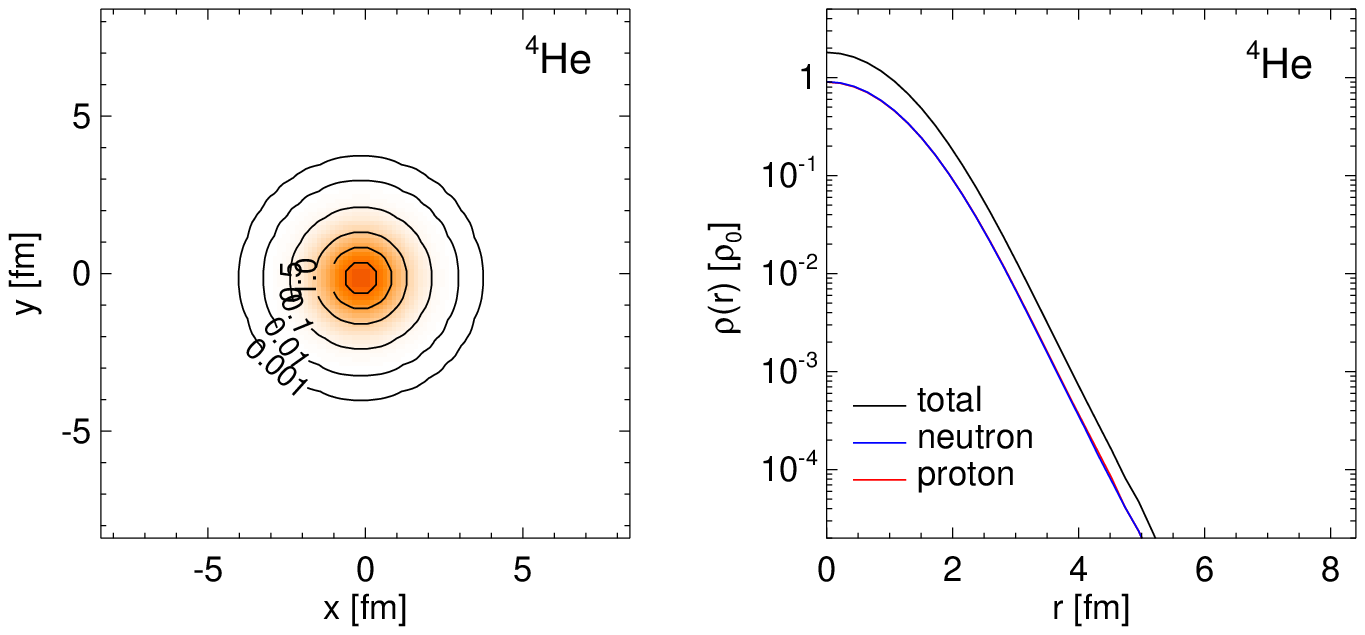}\hfil
  \includegraphics[angle=0,width=0.48\textwidth]{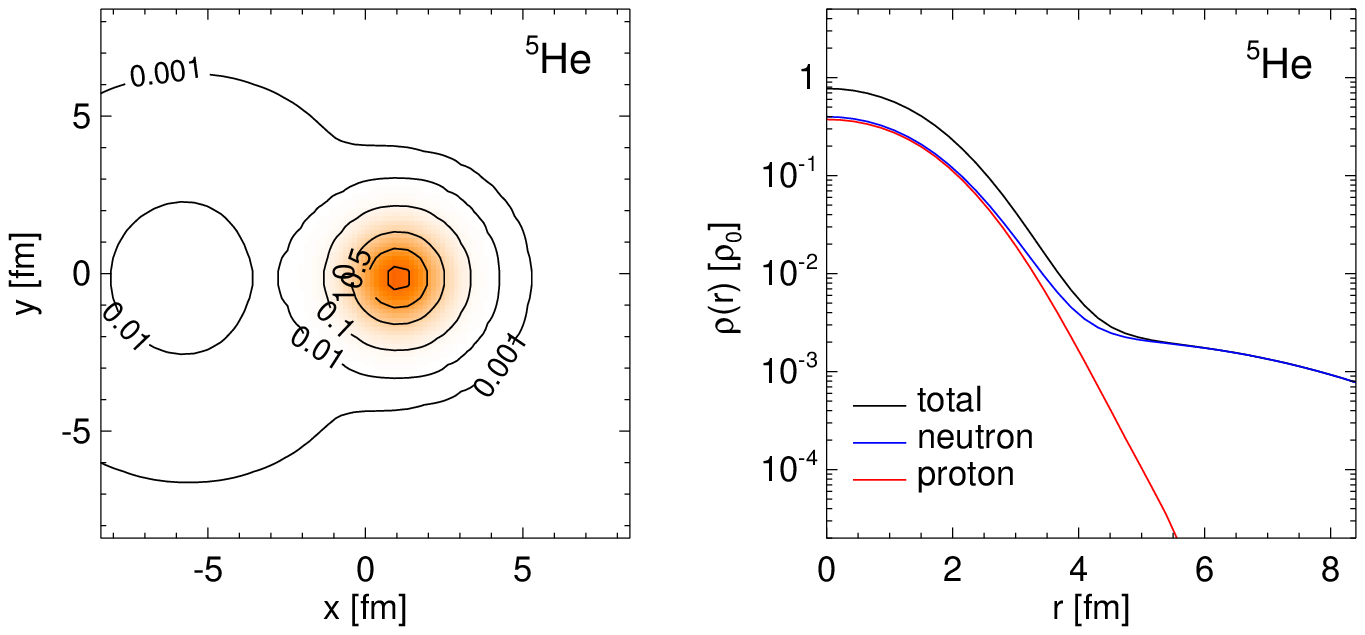}\\[-6ex]
  \includegraphics[angle=0,width=0.48\textwidth]{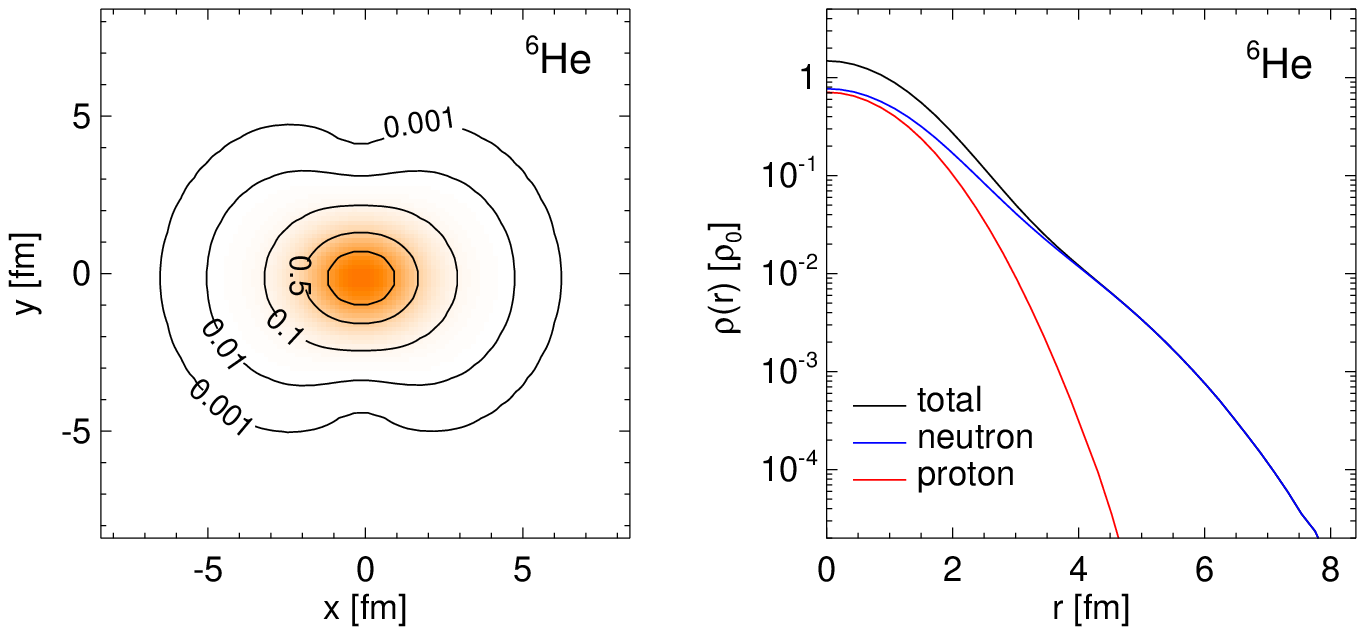}\hfil
  \includegraphics[angle=0,width=0.48\textwidth]{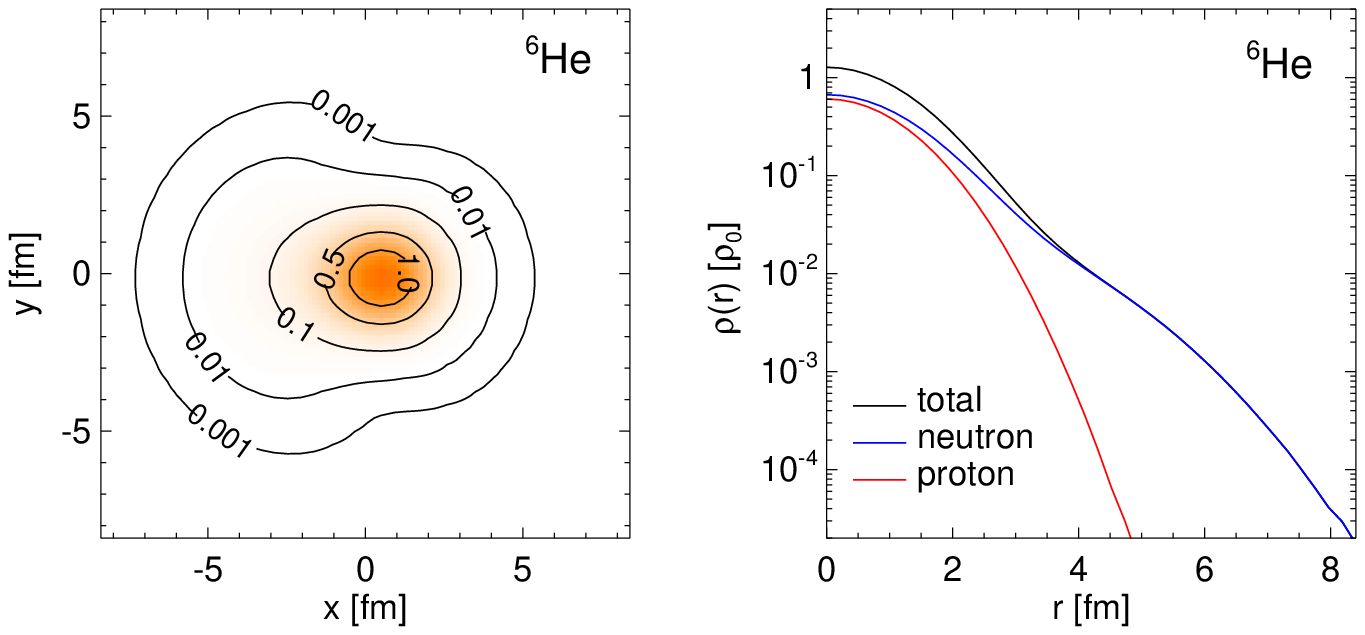}
\\[-6ex]
  \includegraphics[angle=0,width=0.48\textwidth]{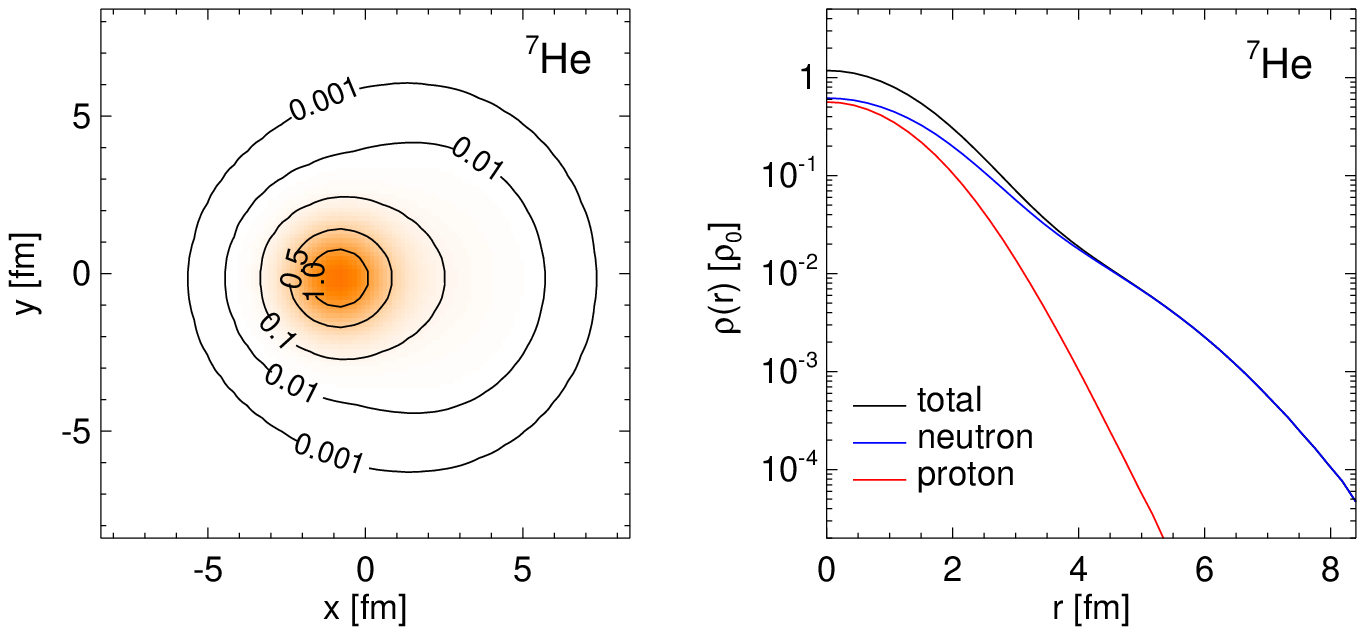}\hfil
  \includegraphics[angle=0,width=0.48\textwidth]{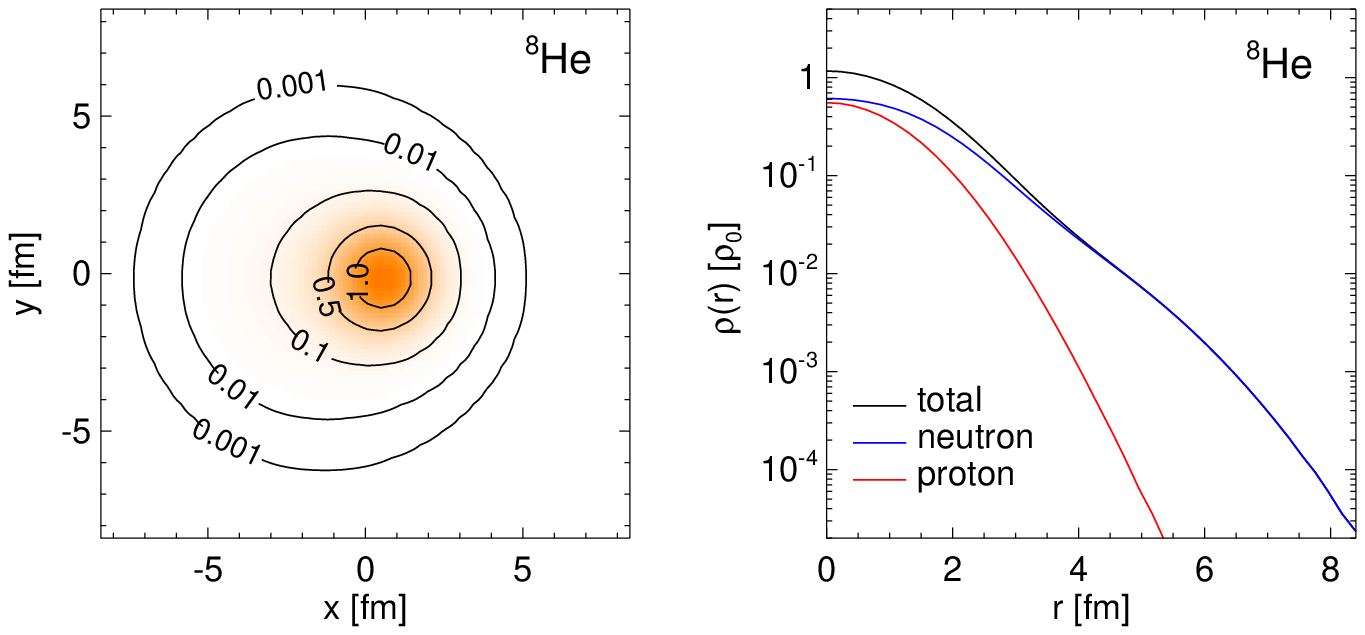}\\
  \vspace{-10ex}
  \caption{L.h.s.: intrinsic shapes corresponding to VAP
    minima. Cuts through nucleon density at $z=0$. R.h.s.: proton,
    neutron and nucleon densities as a function of radius.}
  \label{fig:heisotopes}
\end{figure}

To calculate the properties of the He isotopes we performed FMD
calculations with two Gaussians per single-particle state using the
electric dipole and the mass quadrupole moment as constraints for
variation after projection (VAP) and multiconfiguration (Multiconf)
calculations. In Table~\ref{tab:heisotopes} the calculated binding
energies and radii are shown together with the experimental values. In
Figure~\ref{fig:heisotopes} the intrinsic configurations corresponding
to the VAP minima are displayed. With the two Gaussians the FMD wave
function is able to describe the neutron halo around the
$\alpha$-core. Using only a single Gaussian an imperfect halo together
with a distorted $\alpha$-core is observed. Compared to the
calculation using a single Gaussian the neutron-rich He isotopes gain
about 3-5~MeV in the PAV case.  For the \chemical{6}{He} nucleus we
find a quadrupole deformed VAP minimum where the two halo neutrons are
sitting on opposite sides of the core and a more bound (by about
1.1~MeV) dipole deformed VAP minimum where the neutrons are located on
the same side of the core (see middle row of
Figure~\ref{fig:heisotopes}). For the other isotopes we only find
minima with respect to the dipole deformation. The multiconfiguration
calculations that include shapes of different electric dipole moments
reproduce the experimental staggering of the binding energies.  The
large measured matter radii of \chemical{6}{He} and \chemical{8}{He}
are also well described by the multiconfiguration calculations.  We
also observe a significant increase in the charge radii of the neutron
rich He isotopes compared to \chemical{4}{He}.  This is explained by
the zero-point motion of the more or less unchanged $\alpha$-core with
respect to the center of mass of the nucleus.

We conclude that the zero-point motion of paired neutrons in a
soft-dipole mode seen in VAP and multiconfiguration mixing
calculations is an essential ingredient for the understanding of
neutron-rich He isotopes and probably also for other neutron halo
nuclei.

\end{document}